\documentclass{article}
\usepackage[utf8]{inputenc}
\usepackage{amsmath,amssymb,amsfonts,stmaryrd}
\usepackage{physics}
\usepackage{dsfont}
\usepackage{graphicx}
\usepackage{xcolor}
\usepackage{graphicx}
\usepackage{cancel}
\usepackage{hyperref}
\usepackage{bm} %allows for bold math type
\usepackage{subfigure} %allows for 2x2 figures
\usepackage{environ}
\usepackage{url}
\usepackage{hyperref}
\usepackage[margin=0.75in]{geometry}

\newcommand{\Z}{{\mathbb Z}}

\NewEnviron{eqs}{%
\begin{equation}\begin{split}
    \BODY
\end{split}\end{equation}
}

\title{Higher-Form Anomalies Imply Intrinsic Long-Range Entanglement}
\author{Po-Shen Hsin$^{1}$, Ryohei Kobayashi${}^2$, Abhinav Prem${}^2$}

\begin{document}

  \bigskip

\maketitle
\begin{center}
${}^1$Department of Mathematics, King’s College London, Strand, London WC2R 2LS, UK

${}^2$School of Natural Sciences, Institute for Advanced Study, Princeton, NJ 08540, USA
\end{center}

\bigskip\bigskip
\begin{abstract}
We show that generic quantum many-body states which respect an anomalous finite higher-form symmetry have an exponentially small overlap with any short-range entangled (SRE) state. Hence, anomalies of higher-form symmetries enforce \textit{intrinsic} long-range entanglement, which is in contrast with anomalies of ordinary (0-form) symmetries which are compatible with symmetric SRE states (specifically, symmetric cat states). As an application, we show that the anomalies of strong higher-form symmetries provide a diagnostic for mixed-state topological order in $d \geq 2$ spatial dimensions. We also identify a new (3+1)D intrinsic mixed-state topological order that does not obey remote-detectability by local decoherence of the (3+1)D Toric Code with fermionic loop excitations.
This breakdown of remote detectability, as encoded in anomalies of strong higher-form symmetries, provides a partial characterization of intrinsically mixed-state topological order.
\medskip
\noindent
\end{abstract}

\bigskip \bigskip \bigskip
 
% \date{\today}

\bigskip

% \eject

\tableofcontents

\unitlength = .8mm

\setcounter{tocdepth}{3}

\bigskip

%%%%%%%%%%%%%%%%%%%%%%%%%%%%%%%%%%%%%%%%%%%%%%%%%%%%%
%%%%%%%%%%%%%%%%%%%%%%%%%%%%%%%%%%%%%%%%%%%%%%%%%%%%%

\section{Introduction}
\label{sec:intro}

Symmetries and their associated anomalies provide a powerful organizing framework for understanding the landscape of quantum many-body phases in strongly interacting local systems. Initially discovered in the context of quantum field theory~\cite{adler1969,bell1969,thooft}, the role of anomalies in placing non-perturbative constraints on strongly coupled many-body systems is now widely appreciated across the high-energy, condensed matter, and quantum information communities. In their modern formulation, quantum anomalies arise when global symmetries cannot be consistently gauged, revealing an inherent obstruction to localizing the symmetry action i.e., making the symmetry act ``on-site" (see e.g., Ref.~\cite{shirley2025dis} for a recent discussion). Consequently, any state invariant under an anomalous symmetry cannot be ``trivial" i.e., it cannot be short-range entangled~\cite{chen2010}. For instance, the celebrated Lieb-Schultz-Mattis-Oshikawa-Hastings theorem~\cite{LSM1961,oshikawa2,hastings2004} and its many generalizations~\cite{cheng2016set,cho2017anomaly,Kobayashi2019lsm,else2020lsm,prem2020lsm,seiberg2022lsm} encode non-trivial constraints on the low-energy physics of local lattice systems which stem from mixed anomalies between spatial and internal symmetries. Anomalies also place strong constraints on e.g., deconfinement in gauge theories~\cite{shimizu2018,hsin2020se,seifnashri2021sym} and the nontrivial boundary states of symmetry protected topological phases~\cite{chen2013SPT,else2014,tiwari2018}.

Crucially, anomalies are properties of the symmetry action on the Hilbert space~\cite{Kapustin:2024rrm} and not of any particular Hamiltonian or state: this endows them with a topological robustness and makes them particularly well-suited for inferring the universal features of many-body states that are invariant under smooth local deformations. In other words, anomalies provide a promising conceptual framework for classifying quantum many-body phases. In the context of ground states of gapped local Hamiltonians in (2+1)D, the extension of this framework to higher-form symmetries~\cite{ggs2015, feng2025higherformanomalies} has proven particularly fruitful: the anomalies of 1-form symmetries, which act on extended loop-like objects rather than point-like operators, encode the nontrivial braiding and statistics of anyons in Abelian topological orders. As such, distinct Abelian topological orders can be fully characterized in terms of their anomalous 1-form symmetries. Concurrently, long-range entanglement (LRE) is also a hallmark of such topologically ordered states~\cite{wenreview}: in contrast to short-range entangled (SRE) states, which can be transformed into product states via finite-depth local unitary circuits, LRE states possess intrinsic entanglement that is locally robust. A key recent insight from Ref.~\cite{li2024anyon} is that this LRE is in fact a consequence of anomalous 1-form symmetries. Moreover, rapid progress in the study of mixed-state phases~\cite{Sohal:2024qvq,Ellison:2024svg,Wang2025intrinsic,zhang2024, Lee2023criticality, Fan2024diagnostics, bao2023mixed, Chen2024separability} has unveiled how the anomalies of ``strong" and ``weak" 1-form symmetries (we review these concepts in Sec.~\ref{sec:mixed}) can lead to patterns of many-body entanglement in mixed-states beyond those expected in ground states of gapped local Hamiltonians. Despite these developments, it remains unclear whether anomalies of $p$-form symmetries ($p\geq 1$) necessarily imply intrinsic long-range entanglement on quantum many-body states in arbitrary spatial dimensions and, if so, whether these anomalies can shed light on the classification of mixed-state topological order in $d>2$ spatial dimensions, which remains far from well understood.

In this work, we study the general consequence of 't Hooft anomalies on the entanglement structure of strongly interacting quantum many-body systems in arbitrary spatial dimensions. Specifically, we consider quantum many-body systems in $d$ spatial dimensions which respect an anomalous finite higher-form symmetry, i.e., a $p$-form symmetry with $p\ge 1$. Following Ref.~\cite{kobayashi2024universal}~\footnote{Anomalies of higher-form symmetries are also studied in e.g. Refs.~\cite{ggs2015,Wen:2018zux,Hsin:2018vcg,Barkeshli:2022edm, feng2025higherformanomalies}.}, we characterize the anomalies of finite higher-form symmetries for symmetric states $\ket{\Psi}$ via an invariant dubbed the ``generalized statistics." The generalized statistics is a Berry phase invariant that is obtained by evaluating a sequence of unitary symmetry operators on the state $\ket{\Psi}$. A well-known example of such an invariant is the self statistics of point-like Abelian anyons in (2+1)D topologically ordered states, which can be characterized through the so-called ``T-junction" invariant~\cite{Levin2003Fermions}. Heuristically, the generalized statistics invariant is an extension of the T-junction invariant for anyons to invariants for symmetry defects created at generic symmetric states, gapped or gapless.

Our main result is that any symmetric state $\ket{\Psi}$ with anomalous finite higher-form symmetry hosts ``intrinsic'' bipartite long-range entanglement, in the sense that the overlap of the state $\ket{\Psi}$ with \textit{any} short-range entangled (SRE) state is exponentially decaying with respect to system size. This statement generalizes the result of Ref.~\cite{li2024anyon} which showed that the mutual braiding or self-statistics of point-like anyons in (2+1)D implies a similar constraint for the ground states of specific lattice models. We remark that \textit{intrinsic} long-range entanglement does not generally follow from a global (0-form) symmetry, since anomalies of (invertible) 0-form symmetries can be saturated by symmetric cat states, which have finite overlap with SRE states. As such, anomalies of higher-form symmetries can be viewed as placing considerably stronger constraints on the entanglement structure of quantum many-body states than 0-form symmetries.

As an application of our result, we consider mixed-states that arise from subjecting topologically ordered pure-states to local decoherence channels and, in the spirit of Refs.~\cite{Lessa:2024wcw,Hsin:2023jqa,Zhou:2023icb,Zang:2023qou,Wang:2024vjl,Sohal:2024qvq,Ellison:2024svg,chirame2024to,xu2025avg,zhang2024,Zhou:2025bal,lessa2025higher,zhou2025finiteT}, utilize anomalies to characterize nontrivial mixed-state phases.\footnote{See also \cite{sun2025anomalousMPO} for implications of anomalous non-invertible symmetries in mixed states.} In particular, as an explicit example we consider the (3+1)D $\Z_2$ Toric Code model with a fermionic loop excitation and subject its ground state to a local noise channel. We show that the decohered mixed-state $\rho$ is protected by a subtle `t Hooft anomaly of a 1-form symmetry and that invariance under this anomalous finite higher-form symmetry implies intrinsic long-range entanglement i.e., the fidelity between $\rho$ and \textit{any} SRE mixed-state $\sigma_{\text{SRE}}$ decays exponentially with system size. Consequently, we argue that this state provides an instance of a new intrinsically mixed-state topological order in (3+1)D, which has no pure-state counterpart. While an explicit model for this new mixed topological order is constructed in (3+1)D, our results on intrinsic long-range entanglement hold for any finite anomalous $p$-form symmetry with $p\ge 1$ in any spatial dimensions. As we briefly comment on below, intrinsic long-range entanglement is a stronger constraint enforced by finite higher-form symmetries than constraints that have previously appeared in the literature (such as constraints based on separability~\cite{lessa2024anomaly,chen2023separable,chen2023separable2}) and provides a valuable resource for distinguishing mixed-state phases of matter.

Finally, we discuss how the breakdown of remote detectability provides a sufficient condition for distinguishing intrinsically mixed-state topological order from pure-state topological order in arbitrary spatial dimensions. Remote detectability is an axiom of pure state topological order, as first proposed in \cite{Kong:2014qka,kong2015boundarybulkrelationtopologicalorders,Kong_2017}, see also \cite{Kitaev_2012,Levin_2013} for earlier analysis. It can be formulated mathematically as follows: \cite{Johnson_Freyd_2022}
\begin{equation}
\textbf{Remote Detectability:\quad }\text{The algebra of operators in a topological order has trivial center.
}    
\end{equation}
Physically, it means that for every nontrivial topological operator supported on $n$-dimensional submanifold $M$, such that $M$ forms Hopf link with a $(D-n-2)$-dimensional submanifold, there must exist a nontrivial topological operator supported on the $(D-n-2)$-dimensional submanifold such that these two operators have nontrivial correlation function, i.e. we cannot move the two operators apart to unlink them. Remote detectability generalizes the modularity condition in (2+1)D anyon theories and is expected to hold for topological orders (modulo invertible topological orders) occurring in ground states of gapped local Hamiltonians. Specifically, no topological excitations that are not condensation descendants \cite{Gaiotto:2019xmp} in the theory should be ``transparent'', i.e. undetectable via braiding with other topological excitations (including itself). 
This can be translated to the property that the higher-form symmetries corresponding to the full topological excitations are anomalous, i.e. we cannot condense all excitations, with the anomaly captured by the remote detection braiding correlation functions.

As initially shown in Refs.~\cite{Sohal:2024qvq,Ellison:2024svg}, (2+1)D mixed-states can host strong higher-form symmetries which violate remote detectability and are algebraically characterized in terms of pre-modular (or nonmodular) tensor categories. This is widely believed to be impossible in the ground states of gapped local Hamiltonians; hence, density matrices which host non-modular strong 1-form symmetries were dubbed as carrying ``intrinsically mixed-state topological order (imTO)." We will adopt this definition of imTO in this paper and argue that the violation of remote detectability for the strong higher-form symmetries of (3+1)D mixed-states provides a sufficient condition for imTO.

This paper is organized as follows: in Sec.~\ref{subsec:review}, we first review the generalized statistics invariant that characterizes the anomalies of higher-form symmetries. In Sec.~\ref{subsec:intrinsic}, we show the intrinsic long-range entanglement of symmetric states carrying a nontrivial generalized statistics invariant for finite higher-form symmetries. We apply this result to mixed-state phases in Sec.~\ref{sec:mixed}, where we discuss how anomalies of finite higher-form symmetries can be used to characterize mixed-state topological orders. As a concrete example, we subject the (3+1)D $\Z_2$ Toric Code with fermionic loop excitations to local decoherence and show that the resulting decohered phase realizes an intrinsically mixed-state topological order in (3+1)D. We conclude with a discussion of open questions and future directions in Sec.~\ref{sec:cncls}, with various technical details relegated to the Appendices: Appendix~\ref{app:statistics} discusses the fermionic loop statistics, while Appendix~\ref{app:average} discusses the implications of a \textit{weak} fermionic loop symmetry on the entanglement entropy.

%%%%%%%%%%%%%%%%%%%%%%%%%%%%%%%%%%%%%%%%%%%%%%%%%%%%%

\subsection{Relation to previous works}

We note that a number of recent works have discussed the anomalies of higher-form symmetries and their constraints on the entanglement structure of pure or mixed-states. Here, we clarify the relation between our results and those established in these other works.

\begin{itemize}
\item Ref.~\cite{li2024anyon} focuses on a number of topologically ordered lattice models in (2+1)D that host anomalous finite 1-form symmetries (generated by string operators of anyons) and derives the intrinsic long-range entanglement of states within the ground-state subspace as enforced by the nontrivial anomalies. Our result in Sec.~\ref{sec:LRE} can be regarded as a generalization of this result to generic $p$-form symmetries (with $p\ge 1$) and in arbitrary spatial dimensions $d\ge 2$, without reference to specific lattice models.

\item Ref.~\cite{lessa2025higher} also primarily focuses on (2+1)D systems with anomalous finite 1-form symmetries and shows that the anomalies enforce long-range entanglement for mixed-states. Long-range entanglement of a mixed-state $\rho$ is defined as the absence of two-way connectivity between $\rho$ and any SRE mixed-state $\sigma_{\text{SRE}}$ via stochastic finite depth local channels in Ref.~\cite{lessa2025higher}. In contrast, we show that the strong anomalous higher-form symmetries enforce the exponentially decaying fidelity $\mathcal{F}(\rho,\sigma_{\text{SRE}})=O(L^{-\infty})$ between $\rho$ and any SRE mixed-state as a function of system size $L$, which we take as our definition of \textit{intrinsic} long-range entanglement following Ref.~\cite{li2024anyon}. We emphasize that the latter condition is stronger than the former: first, since neither quasi-local channels nor stochastic local channels can induce intrinsic LRE in short-depth~\cite{lessa2025higher,zhou2025finiteT,piroli2020} (i.e., in depth that scales polylogarithmically with system-size), a state with intrinsic LRE cannot be two-way channel connected to an SRE state. Hence, our constraint implies the absence of two-way channel connectivity. On the other hand, the converse is false i.e., absence of two-way channel connectivity does not imply intrinsic LRE. Consider e.g., a density matrix $\rho = p\rho_{\text{LRE}} + (1-p)\rho'_{\text{SRE}}$ which is a superposition of an intrinsic LRE and an SRE state. Clearly, this cannot be prepared from an SRE state via stochastic quasi-local channels in short-depth but $\rho$ still violates the fidelity relation $\mathcal{F}(\rho,\sigma_{\text{SRE}})\neq O(L^{-\infty})$ i.e., it does not have intrinsic LRE. Thus, intrinsic LRE due to anomalous higher-form symmetries is a stronger constraint than the two-way channel condition discussed in previous works.

We also note that our derivation of intrinsic long-range entanglement is based on the microscopic characterization of anomalies developed in Ref.~\cite{kobayashi2024universal}, which is shown to be an invariant of generic symmetric states, gapped or gapless. Ref.~\cite{lessa2025higher} also uses a specific quantity to characterize the anomalies of 1-form symmetries, and it would be interesting to see if this characterization is also invariant against local perturbations. Ref.~\cite{lessa2025higher} also comments on a subset of higher-form anomalies in generic spacetime dimensions and conjectures that a subset of higher-form anomalies leads to \textit{multipartite} entanglement. In contrast, we consider generic higher-form anomalies characterized by the generalized statistics developed in Ref.~\cite{kobayashi2024universal} and show that they imply intrinsic long-range bipartite entanglement. This includes generic higher-form anomalies whose topological responses are expressed in terms of Steenrod or Pontryagin operations of background gauge fields. It would be interesting to study the implications on the multipartite entanglement enforced by the generalized statistics of strong higher-form symmetries.

\item Refs.~\cite{Sohal:2024qvq,Ellison:2024svg} propose and obtain a family of intrinsically mixed-state topological orders in (2+1)D by subjecting (2+1)D pure state topological orders to local noise channels, which lead to the incoherent proliferation of certain anyons. These papers conjecture a classification of intrinsically mixed-state topological order in (2+1)D in terms of non-modular strong 1-form symmetries and their anomalies. We note that all examples of intrinsically mixed-state topological phases that have appeared in the literature thus far are enforced by anomalous symmetries generated by topological line operators. We provide the first example where the intrinsically mixed-state topological order in the (3+1)D $\Z_2$ Toric Code results from loop excitations with nontrivial self-statistics, instead of point-like quasiparticles. The topological order of this mixed state, as described by the strong symmetry of the state, is a theory of a single fermionic loop which violates the remote detectability condition that is expected to hold for the ground states of gapped local Hamiltonians~\cite{PhysRevX.8.021074,Johnson_Freyd_2022}. Based on this, we conjecture that the breakdown of remote detectability as encoded in the anomalies of strong higher-form symmetries provides a sufficient condition for distinguishing intrinsically mixed-state topological order from pure-state topological order.

\end{itemize}

%%%%%%%%%%%%%%%%%%%%%%%%%%%%%%%%%%%%%%%%%%%%%%%%%%%%%
%%%%%%%%%%%%%%%%%%%%%%%%%%%%%%%%%%%%%%%%%%%%%%%%%%%%%

\section{Intrinsic Long-Range Entanglement}
\label{sec:LRE}

Here, we consider arbitrary finite $p$-form (with $p \geq 1$) symmetries in general spacetime dimensions and show that their anomalies, as defined via the generalized statistics introduced in Ref.~\cite{kobayashi2024universal}, implies intrinsic long-range entanglement.

\subsection{Review: Higher-form anomalies and generalized statistics}
\label{subsec:review}

We first review the generic microscopic characterization of 't Hooft anomalies for finite $p$-form symmetries, following Ref.~\cite{kobayashi2024universal}. We consider a state $\ket{\Psi}$ with a finite $G$ $p$-form symmetry on a tensor product Hilbert space.
The symmetry is generated by topological operators $U_g(\Sigma)$ (labeled by $g\in G$) that are supported on a $(d-p)$-dimensional manifold $\Sigma$ embedded in $d$-dimensional space. When $\Sigma$ is closed and contractible, $U_g(\Sigma)$ leaves the state $\ket{\Psi}$ invariant up to an overall U(1) phase: $U_g(\Sigma)\ket{\Psi}\propto\ket{\Psi}$.  
We will also assume that $U_g(\Sigma)$ is given by a finite-depth unitary circuit, such that one can define $U_g(\Sigma)$ on either closed or open manifolds $\Sigma$. When $\Sigma$ is not closed, $U_g(\Sigma)$ does not leave $\ket{\Psi}$ invariant at the boundary and instead creates nontrivial symmetry defects at the boundary $\partial\Sigma$. We will consider only higher-form symmetries with $p\ge 1$, so $G$ is necessarily an Abelian group (we will not consider non-invertible higher-form symmetries here). 

We then consider a set of states with symmetry defects $\{\ket{a}\}$ containing configurations of possible symmetry defects, where $a$ labels a configuration of symmetry defects as we describe below. These states can be obtained by applying the symmetry operators $U_g(\Sigma)$ with open manifolds $\Sigma$ on the input symmetric state $\ket{\Psi}$, and we give the construction of the set $\{\ket{a}\}$ below in detail.

\begin{figure*}[t]
    \centering
    \subfigure[(1+1)D]{\raisebox{2cm}{\includegraphics[scale=0.7]{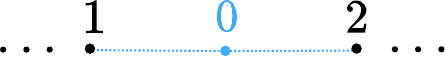}}\label{fig: complex (a)}}
    \hspace{0.05\linewidth}
    \subfigure[(2+1)D]{\includegraphics[scale=0.7]{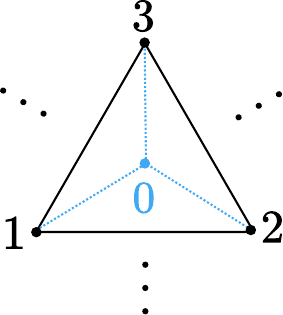}\label{fig: complex (b)}}
    \hspace{0.05\linewidth}
    \subfigure[(3+1)D]{\includegraphics[scale=0.7]{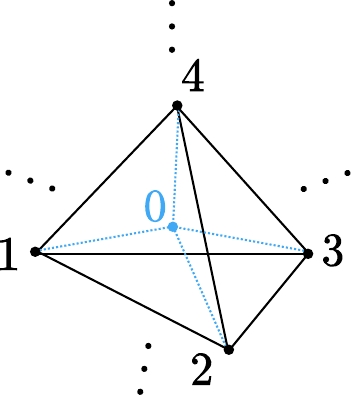}\label{fig: complex (c)}}
    \caption{The simplicial complex $X$ embedded in (a) 1d, (b) 2d, and (c) 3d space. This is a $d$-simplex with a center vertex $0$, dividing a single $d$-simplex into $d+1$ of them. This is regarded as a minimal triangulation of a $d$-sphere, once we identify infinity as a single point. Symmetry defects are supported at the $(d-p-1)$-simplices of the simplicial complex.
    }
    \label{fig: complex}
\end{figure*}

Let $\mathcal{A}$ be a finite set of configurations of symmetry defects, such that $a\in\mathcal{A}$. To fix the definition of $\mathcal{A}$, we introduce a simplicial complex $X$ embedded in the space, which we choose to be a minimal triangulation of a $d$-dimensional sphere (see Figure~\ref{fig: complex}).
We consider the $p$-form symmetry operators $U_g(\Sigma)$ supported at a $(d-p)$-simplex $\Sigma$ of the simplicial complex $X$. The symmetry operator is hence labeled by a pair $(g,\Sigma)$, which is regarded as a generator of the $(d-p)$-chain $C_{d-p}(X,G)$ (see Appendix \ref{app:complex} for a review of simplicial complexes).

When the open symmetry operator labeled by $s\in C_{d-p}(X,G)$ acts on a quantum state, it creates a symmetry defect on its boundary $a=\partial s$. Therefore the symmetry defects are labeled by boundaries $\mathcal{A} = B_{d-p-1}(X,G)$ of the chain complex, meaning that the symmetry defects are created at the $(d-p-1)$-simplices that are boundaries of symmetry operators. Each state $\ket{a}$ with $a\in\mathcal{A}$ can be obtained (up to an overall phase) by applying a sequence of symmetry operators $U_g(\Sigma)$ on $\ket{\Psi}$. The action of symmetry operators labeled by $s\in C_{p-1}(X,G)$ is expressed in the form of
\begin{align}
    U(s)\ket{a} = \exp[i\theta(s,a)]\ket{a+\partial s}
    \label{eq:def of theta}
\end{align}
with $\theta(s,a)$ an overall phase.

As an illustration, consider a (3+1)D system with a $1$-form symmetry ($p=1$). Here, the chain $s \in C_2(X,G)$ represents a collection of oriented plaquettes labeled by group elements $g\in G$, while its boundary $a=\partial s \in B_1(X,G)$ corresponds to a set of closed loops on edges that correspond to the loci of symmetry defects. Acting with the operator $U(s)$ thus creates loop excitations along these edges, consistent with the geometric picture shown in Figure~\ref{fig:24step} (a).
In this way, the chain complex $(C_\bullet(X,G), \partial)$ provides the natural algebraic framework for expressing the action of higher-form symmetries on a lattice.

We emphasize that to obtain a set of such symmetry operators $U_g(\Sigma)$ on open manifolds, one generally needs to make a specific choice of the termination of the symmetry operator $U_g(\Sigma)$ at the boundary $\partial \Sigma$. For instance, denoting the order of $g\in G$ by $|g|$, the symmetry operator with boundaries must be chosen so that $U_g(\Sigma)^{|g|}$ preserves the state $\ket{\Psi}$ up to an overall phase. This ensures that symmetry defects follow the same group algebra as the symmetries under fusion, so that the defects are labeled by $\mathcal{A} = B_{d-p-1}(X,G)$ with coefficient in $G$. Then, the generalized statistics has the form
\begin{align}
    e^{i\Theta} = \bra{\Psi}\prod_j U^{\pm 1}(s_j)\ket{\Psi}~,
    \label{eq:statistics}
\end{align}
which evaluates the expectation value of the finite sequence of unitary operators $U^{\pm 1}(s_j)$, where each $s_j$ is a generator of $C_{d-p}(X,G)$. 
The invariant $\Theta$ is a Berry phase given by the sum of $\theta(s,a)$. $\Theta$ is in general not invariant under local perturbations of the state, but becomes a quantized invariant of a symmetric state given a deliberate choice of the unitary sequence. The choice of the signs $\pm 1$ depends on each operator $U(s_j)$, which is again carefully chosen so that the Berry phase becomes an invariant.
When it becomes an invariant, $\Theta$ is referred to as the ``generalized statistics", since it generalizes the notion of microscopic self-statistics or mutual statistics of point-like anyons to excitations or symmetry defects with arbitrary spatial extent (such as loop-like excitations).\footnote{If we restrict ourselves to gapped phases, symmetry defects of higher-form symmetry are identified as the excitations including anyons.} Ref.~\cite{kobayashi2024universal} derived the necessary and sufficient conditions for the unitary sequence to define a robust quantized invariant.

When the $G$ symmetry is gauged, Each symmetry operator $U(s)$ is regarded as a product of Gauss law operators.
The generalized statistics $\Theta$ can also be regarded as the product of Gauss law operators for the $G$ symmetry. 
If the generalized statistics $e^{i\Theta}$ is nontrivial, it implies one cannot solve the Gauss law constraints $U(s)=1$, thus defines an obstruction to gauging the $G$ symmetry. Therefore, the generalized statistics gives the microscopic definition of the 't Hooft anomaly of a $G$ symmetry. Ref.~\cite{kobayashi2024universal} conjectured that such an invariant characterizes all 't Hooft anomalies of finite higher-form symmetries in bosonic theories. \footnote{See Ref.~\cite{feng2025anyonicmembranes} for supporting evidence of this conjecture in high spacetime dimensions.}

In continuum QFT, 't Hooft anomaly leads to the dynamical consequence which forbids a trivial symmetric gapped state. Accordingly, the
generalized statistics $\Theta$ forbids short-range entanglement for any invariant ground state: if $\ket{\Psi}$ respects an anomalous finite $p$-form symmetry with a nontrivial invariant $\Theta\neq 0$ mod $2\pi$, it cannot be an SRE state $\ket{\Psi}\neq V_t\ket{0}$ i.e., it cannot be connected via any local unitary circuit $V_t$ with constant depth $t$ to a trivial product state $\ket{0}$~\cite{kobayashi2024universal}. Below, we briefly review a proof of this statement using the example of a $\Z_2$ 1-form symmetry in (3+1)D.

\subsubsection{Example: Fermionic loops in (3+1)D}
\label{subsec:floop}
\begin{figure}[htb]
\centering
\includegraphics[width=\textwidth]{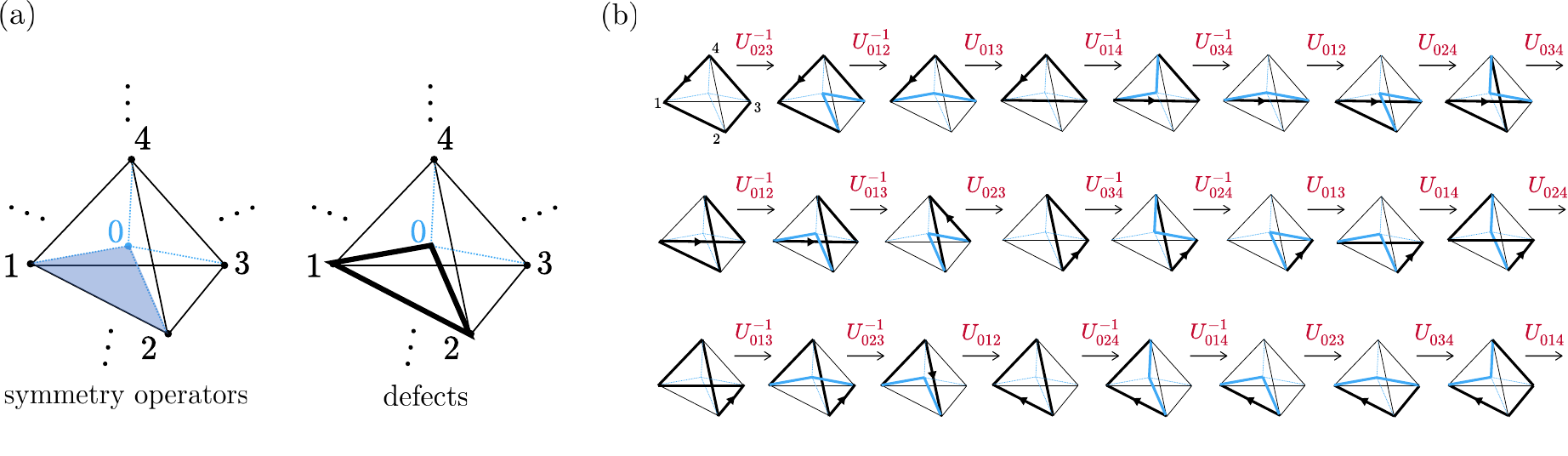}
\caption{(a): 1-form symmetry in (3+1)D. The symmetry operators are supported at 2-simplices, and the symmetry defects are supported at their boundaries. (b): The invariant for the $\Z_2$ 1-form symmetry in (3+1)D is defined by a sequence of 24 symmetry operators $U_{0jk}$ evaluated on a symmetric gapped state $\ket{\Psi}$. The thick blue or black lines denote the configurations of symmetry defects during the 24-step process. The arrow along the thick line indicates that the initial and final loops of defects are reversed. Therefore this 24-step process is sometimes called the loop-flipping process.}
\label{fig:24step}
\end{figure}

Let us consider a (3+1)D gapped phase with a $\Z_2$ 1-form symmetry. In this case, there is a single $\Z_2$ invariant associated with the self-statistics of the loop excitations (loop-like symmetry defects). The invariant is defined as the expectation value of 
    \begin{eqs}
        U_\Theta := ~& U_{014} U_{034} U_{023} U_{014}^{-1} U_{024}^{-1} U_{012} U_{023}^{-1} U_{013}^{-1} \\
        \times & U_{024} U_{014} U_{013} U_{024}^{-1} U_{034}^{-1} U_{023} U_{013}^{-1} U_{012}^{-1} \\
        \times & U_{034} U_{024} U_{012} U_{034}^{-1} U_{014}^{-1} U_{013} U_{012}^{-1} U_{023}^{-1}~,
        \label{eq: 24-step process}
    \end{eqs}
where $U_{0jk}$ is the $\Z_2$ symmetry defect on a triangle, as shown in Figure~\ref{fig:24step} (b). On a symmetric state $\ket{\Psi}$, the anomalous $\Z_2$ 1-form symmetry has $U_\Theta=-1$, while a non-anomalous symmetry has $U_{\Theta}=+1$.

A loop excitation or defect associated with this nontrivial $\Z_2$ invariant $U_\Theta=-1$ is called a fermionic loop.
Here, we review the result of Ref.~\cite{kobayashi2024universal} that $\ket{\Psi}$ is not SRE when $\ket{\Psi}$ has the $\Z_2$ 1-form symmetry with the nontrivial $\Z_2$ invariant. If $\ket{\Psi}$ has such a 1-form symmetry, the state $V^\dagger_t\ket{\Psi}$ produced via a finite depth $t$ local unitary circuit $V_t$ also has an anomalous $\Z_2$ 1-form symmetry, which is generated by the dressed symmetry operators $V^\dagger_t U_{0jk} V_t$. Therefore, to show that $\ket{\Psi}$ cannot be an SRE state $V_t\ket{0}$, it suffices to show that the new state $V^\dagger_t\ket{\Psi}$ with anomalous $\Z_2$ 1-form symmetry cannot be a trivial product state $\ket{0}$. Below, we simply rewrite $V^\dagger_t\ket{\Psi}$ as $\ket{\Psi}$, and symmetry operators $V^\dagger_t U_{0jk} V_t$ as $U_{0jk}$, and show that $\ket{\Psi}$ cannot be a trivial product state.

For later purposes, we show a slightly stronger statement: let us consider a disk region $R$ of linear size $r$ in the 3d space. Here, $r$ is taken large enough to define the generalized statistics invariant \eqref{eq: 24-step process} of $\ket{\Psi}$ by the operators $U_{0jk}$ supported within the region $R$. Then, suppose that $\ket{\Psi}$ has the generalized statistics $U_\Theta= -1$ for the $\Z_2$ 1-form symmetry.
Then, the reduced density matrix $\rho_R = \mathrm{Tr}_{R^c}(\ket{\Psi}\bra{\Psi}) = \sum_k \alpha_k\ket{\Psi_k}\bra{\Psi_k}$ ($R^c$ is the complement of $R$), with the orthogonal states $\ket{\Psi_k}$, satisfies
\begin{align}
    |\ket{\Psi_k} - \ket{0}_R| \ge \epsilon, \quad \text{for any $k$ with $\alpha_k\neq 0$~,}
\end{align}
where $\ket{0}_R$ is any trivial product state at $R$, and $\epsilon$ is a fixed small constant. 

Let us show this statement by contradiction; suppose that $|\ket{\Psi_k} - \ket{0}_R| < O(\epsilon)$. We introduce a shorthand notation $\ket{\psi}\sim \ket{\psi'}$ when $|\ket{\psi}-\ket{\psi'}|<O(\epsilon)$. Since $\ket{\Psi}$ has the $\Z_2$ 1-form symmetry, $U(s)\ket{\Psi}\propto \ket{\Psi}$ when $s$ has the support within $R$ and does not have a boundary. This implies $U(s)\ket{\Psi_k}\propto \ket{\Psi_k}$, so $U(s)\ket{0}_R\sim \ket{0}_R$ up to phase.
This further implies that the state $\ket{0}_R$ also has the $\Z_2$ 1-form symmetry $U(s)$ (up to $< O(\epsilon)$).
Also, if any sequence of symmetry operators $\prod U^{\pm 1}(s)$ (with each $\partial s\neq 0$ in general) within $R$ leaves $\ket{\Psi}$ invariant, $\ket{0}$ is also invariant under this sequence (up to $<O(\epsilon)$).
Therefore, one can define the invariant (up to $< O(\epsilon)$) $U_\Theta$ on $\ket{0}_R$.
Since $U_\Theta \ket{\Psi} = e^{i\Theta}\ket{\Psi}$, we have $U_\Theta \ket{\Psi_k} = -\ket{\Psi_k}$, so the invariant becomes $U_\Theta \ket{0}_R\sim -\ket{0}_R$.

Meanwhile, when the product state $\ket{0}_R$ has the $\Z_2$ 1-form symmetry, we have $U(s)\ket{0}_R \sim \ket{\partial s}_A \otimes \ket{0}_{A^c}$ with the open symmetry operator $U(s)$, where $A$ is the one dimensional locus within $R$ where the excitation created by $U(s)$ is supported, and $A^c$ is its complement in $R$.

The state $\ket{\partial s}_A$ is regarded as a (1+1)D gapped state and thus admits a representation using matrix product states (MPS). Now, each excited state is represented in terms of the reduced 1d MPS state localized at the excitation. One can show that the sequence of unitaries Eq.~\eqref{eq: 24-step process} must evaluate trivially on this reduced 1d state, which was explicitly shown in Sec.~VB of Ref.~\cite{kobayashi2024universal}. For our current purposes, a straightforward generalization of the logic explained in Sec.~VB of Ref.~\cite{kobayashi2024universal} shows that $U_\Theta \ket{0}_R\sim \ket{0}_R$, which can be done simply by replacing the equality $=$ in all equations with $\sim$. This leads to the desired contradiction, so we must have $|\ket{\Psi_k} - \ket{0}_R| \ge \epsilon$.

We note that the above result can be extended to generic finite $p$-form symmetry in $d$ spatial dimensions, with $p\ge 1$. This utilizes the argument of Sec.~VC of Ref.~\cite{kobayashi2024universal}, which argues that the trivial product state $\ket{0}$ with finite $p$-form symmetry must have trivial generalized statistics $U_\Theta=1$. The argument in Ref.~\cite{kobayashi2024universal} for generic $p$-form symmetry again takes the product state $\ket{0}$ with $p$-form symmetry, and considers an excited state which takes the form $U(s)\ket{0} = \ket{\partial s}_A\otimes\ket{0}_{A^c}$ with $\ket{\partial s}_A$ the state at the support of the symmetry defects.
Based on a physically reasonable assumption that $\ket{\partial s}_A$ admits a tensor network representation, it is shown that the generalized statistics invariant $U_\Theta$ has to evaluate trivially on the excited states $U(s)\ket{0} = \ket{\partial s}_A\otimes\ket{0}_{A^c}$.

\subsection{Intrinsic long-range entanglement from higher-form anomalies}
\label{subsec:intrinsic}

Let us consider a (pure) state $\ket{\Psi}$ on a tensor product of $n$ local Hilbert spaces. 
When the system has the linear size $L$, $n$ scales as $n = \Theta(L^d)$.

Suppose that the state $\ket{\Psi}$ has a finite $p$-form $G$ symmetry with $p \ge 1$, with an 't Hooft anomaly of the $p$-form symmetry. The 't Hooft anomaly of the finite $p$-form symmetry is characterized by the generalized statistics Eq.~\eqref{eq:statistics}
\begin{align}
    U_\Theta \ket{\Psi} = e^{i\Theta}\ket{\Psi}
\end{align}
with $\Theta\neq 0$ mod $2\pi$,
where $U_\Theta$ is given by a sequence of symmetry operators on open submanifolds of the space.

Let us consider a generic product state $\ket{0^n}$ and a local finite depth circuit $V_t$ with depth $t$. We introduce a state $\ket{\Psi_t} = V_t\ket{\Psi}$. We want to show that 
\begin{align}
    \mathcal{F}(\ket{0^n}\bra{0^n}, \ket{\Psi_t}\bra{\Psi_t}) =\mathcal{F}(V^\dagger_t\ket{0^n}\bra{0^n}V_t, \ket{\Psi}\bra{\Psi})\le (1-\epsilon)^{\frac{2n}{(\gamma t)^d}}~,
    \label{eq:small_overlap}
\end{align}
with some positive constant $\gamma$. Here, the fidelity is defined as $\mathcal{F}(\rho,\sigma) = \left(\text{Tr}\sqrt{\sqrt{\rho}\sigma\sqrt{\rho}}\right)^2$. The above inequality shows that the anomalous state $\ket{\Psi}$ has exponentially small overlap with any SRE states with circuit depth $t=\Theta(L^\alpha)$ with $\alpha<1$.

\begin{figure}[htb]
\centering
\includegraphics[width=0.35\textwidth]{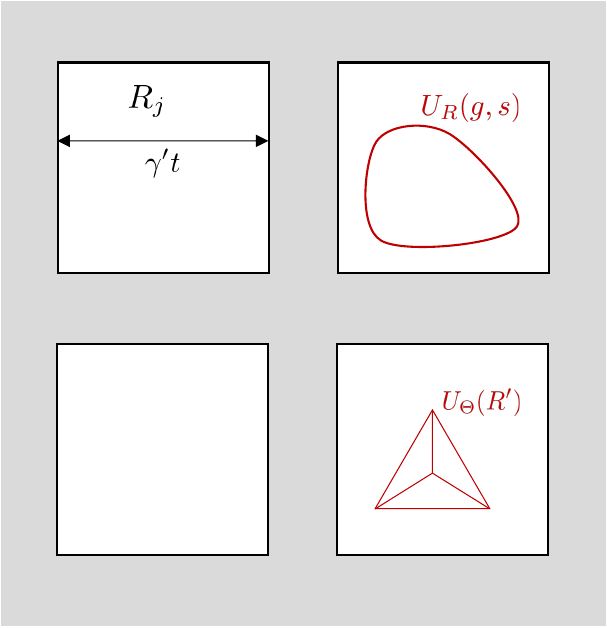}
\caption{The $d$-dimensional space embeds a disjoint set of disks (hypercubes) $\bigcup_j R_j$ with linear size $\gamma't$. The reduced density matrix $\rho_{\{R\}}$ on $\bigcup_j R_j$ has $p$-form symmetry $U_R(g,\Sigma)$ whose support is within a single region $R$. One can then define an invariant $U_\Theta(R)$ by the sequence of symmetry operators within a single region $R$.}
\label{fig:Rj}
\end{figure}

To make the logic of our argument more transparent, we first outline the proof. 
First, we partition the system of linear size $L$ into a disjoint set of regions ${R_j}$ of size proportional to the circuit depth $t$, so that within each region the dressed symmetry operators $V_t U(s) V_t^{\dagger}$ remain well-defined and locally supported (see Figure \ref{fig:Rj}).
The reduced density matrix on the union of these regions, $\rho^{(t)}_{\{R\}}$, inherits a strong $p$-form symmetry because the operators $U(s)$ can be supported entirely within each $R_j$. Inside each region, one can therefore define the generalized statistics operator $U_\Theta(R_j)$ associated with the local symmetry action. The assumption that $|\Psi\rangle$ has an anomalous symmetry means that every such local operator acts as $U_\Theta(R_j)|\Psi\rangle = e^{i\Theta}|\Psi\rangle$ with $e^{i\Theta}\neq 1$.

We then note that for $p\ge1$, the action of the symmetry within each $R_j$ factorizes, so each Schmidt state of a single disk density matrix $\rho^{(t)}_{{R}_m}$ carries the same anomalous symmetry and thus the same nontrivial generalized statistics. By the result of Sec.\ref{subsec:floop}, such a locally symmetric state cannot coincide, even approximately, with any product state; their inner product is bounded by a constant $1-\epsilon<1$. Combining this over all regions yields an exponentially small overlap between $|\Psi\rangle$ and any state that can be prepared from a product state by a finite-depth circuit of depth $t=O(L^{\alpha})$ with $\alpha<1$. Equivalently, the fidelity between $|\Psi\rangle$ and any SRE state decays as $\mathcal{F}\sim(1-\epsilon)^{2n/(\gamma t)^d}$, establishing that an anomalous higher-form symmetry enforces intrinsic long-range entanglement.

\paragraph{Proof of Eq.~\eqref{eq:small_overlap}.}
Now we present an explicit proof for the statement.
Let us consider a disjoint union of disk regions $\bigcup_{j}R_j$ inside the space, where each $R_j$ has linear size $\gamma' t$ with some positive constant $\gamma'$. The number of such regions $R_j$ is given by $n/(\gamma t)^d$, with $\gamma > \gamma'$. 
See Figure~\ref{fig:Rj}. 

The reduced density matrix of $\ket{\Psi_t}$ on $\bigcup_{j}R_j$ is denoted by $\rho^{(t)}_{\{R\}}$. $\rho^{(t)}_{\{R\}}$ is strongly symmetric under the action of dressed $p$-form $G$ symmetry operators $\tilde U_{R_j}(g,\Sigma) = V_tU_{R_j}(g,\Sigma) V_t^\dagger$, where $g\in G$ and $\Sigma$ is the closed $(d-p)$ submanifold support inside $R_j$:
\begin{align}
    \tilde U_{R_j}(g,\Sigma)\rho^{(t)}_{\{R\}} = \lambda(g,\Sigma)\rho^{(t)}_{\{R\}}~,
    \label{eq:strongsymmetry}
\end{align}
with some phase $\lambda(g,\Sigma)$. 

Note that the strong $p$-form symmetry of the reduced density matrix is only present for $p\ge 1$, where one can find the closed symmetry operators supported within the disk region $R$. For 0-form symmetries, in general one cannot find either strong or weak symmetry operators of the reduced density matrix supported within the region $R$, since the symmetry action does not factorize into a tensor product on bipartite Hilbert spaces along the entangling surface. Even when one can find such symmetry operator within $R$, which is possible for an onsite 0-form symmetry, the reduced density matrix is only weakly symmetric in general. This is the case when the symmetry is spontaneously broken and the state $\ket{\Psi}$ is the symmetric cat state.

Due to the Lieb-Robinson bound, note that the linear size of a dressed symmetry operator $V_tU(g,\Sigma) V^\dagger_t$ is increased by $t$ compared with the original size of $U(g,\Sigma)$. 
Suppose now that the size $\gamma't$ of $R_j$ is taken large enough to be able to define the generalized statistics $V_tU_\Theta V^\dagger_t$ of $\ket{\Psi_t}$ within each region $R_j$, using the dressed symmetry operators $\tilde U$ supported within $R_j$. 
The linear size $\gamma't$ of $R_j$ is taken proportional to $t$, so that $V_tU_\Theta V^\dagger_t$ is guaranteed to be supported within the region $R_j$ with the increasing circuit depth.

Then, $\rho^{(t)}_{\{R\}}$ is also strongly symmetric under the action of the generalized statistics $\tilde U_\Theta(R_j) = V_t U_\Theta(R_j)V_t^\dagger$ for each region $R_j$,
\begin{align}
    \tilde U_\Theta(R_j)\rho^{(t)}_{\{R\}} = e^{i\Theta}\rho^{(t)}_{\{R\}}~.
    \label{eq:strongsymmetry Theta}
\end{align}

Let us write $\rho^{(t)}_{\{R\}} = \sum_{k}\alpha_k \ket{\Psi_k}\bra{\Psi_k}$ using the orthogonal basis of $\mathcal{H}_{\{R\}}$. Each $\ket{\Psi_k}$ is symmetric under the operators $\{\tilde U_{R_j}(g,\Sigma), \tilde U_\Theta(R_j)\}$ with the eigenvalues appearing in Eqs.~\eqref{eq:strongsymmetry} and~\eqref{eq:strongsymmetry Theta}. We again note that in the case of 0-form symmetries, $\ket{\Psi_k}$ is generally not symmetric under an operator supported within a single region $R_j$. Meanwhile, it can be symmetric under the product of symmetry operators $\prod_j U_g(R_j)$ which acts globally on all regions $\{R_j\}$, which is possible when the 0-form symmetry is onsite. This underlies the crucial distinction between higher-form and 0-form symmetries. 

One can perform the Schmidt decomposition of each state $\ket{\Psi_k}$, following the tensor product decomposition of the Hilbert space $\mathcal{H}_{\{R\}} = \bigotimes_{j} \mathcal{H}_{R_j}$,
\begin{align}
    \ket{\Psi_k} = \sum_{\{j_m\}}\beta_{\{j_m\};k} \bigotimes_{R_m}\ket{\Psi^{R_m}_{j_m;k}}~,
\end{align}
where $\{\ket{\Psi^{R_m}_{j_m;k}}\}_{\{j_m\}}$ are the orthogonal states of $\mathcal{H}_{R_m}$, where $j_m$ labels the states of $\mathcal{H}_{R_m}$.
Each state $\ket{\Psi^{R_m}_{j_m;k}}$ with nonzero $\beta_{\{j_m\};k}$ is symmetric under the operators 
$\tilde U_{R_m}(g,\Sigma), \tilde U_\Theta(R_m)$, 
\begin{align}
    \tilde U_{R_m}(g,\Sigma) \ket{\Psi^{R_m}_{j_m;k}} = \lambda(g,\Sigma)\ket{\Psi^{R_m}_{j_m;k}}, \quad \tilde U_\Theta(R_m)\ket{\Psi^{R_m}_{j_m;k}} = e^{i\Theta}\ket{\Psi^{R_m}_{j_m;k}}~.
\end{align}

This implies that each state $\ket{\Psi^{R_m}_{j_m;k}}$ has the $p$-form symmetry generated by the dressed operator $\tilde U_{R_m}$, and it carries the nontrivial generalized statistics $e^{i\Theta}$. Now, using the result of Sec.~\ref{subsec:floop} we have
\begin{align}
    \left|\ket{\Psi^{R_m}_{j_m;k}} - \ket{0}_{R_m}\right| \ge \epsilon~,
    \label{eq:away from product state}
\end{align}
with some fixed constant $\epsilon \ll 1$, for any choice of the tensor product state $\ket{0}_{R_m}$.
This implies that the state $\ket{\Psi_k}$ has an exponentially small overlap with the product state,
\begin{align}
    |\langle 0_{\{R\}} \ket{\Psi_k}| \le (1-\epsilon)^{\frac{n}{(\gamma t)^d}}~,
\end{align}
for any choice of the product state $\ket{0_{\{R\}}}$ supported at $\bigcup_j R_j$. Due to the monotonicity of fidelity, we find that
\begin{align}
    \mathcal{F}(\ket{0^n}\bra{0^n}, \ket{\Psi_t}\bra{\Psi_t}) \le \mathcal{F}(\rho^{(t)}_{\{R\}},\ket{0_{\{R\}}}\bra{0_{\{R\}}})\le (1-\epsilon)^{\frac{2n}{(\gamma t)^d}}~.
\end{align}
This is precisely the result we had set out to show.

%%%%%%%%%%%%%%%%%%%%%%%%%%%%%%%%%%%%%%%%%%%%%%%%%%%%%
%%%%%%%%%%%%%%%%%%%%%%%%%%%%%%%%%%%%%%%%%%%%%%%%%%%%%

\section{Applications to Mixed-State Topological Order}
\label{sec:mixed}

We now apply the results of the previous section to mixed-states and show how the anomalies of higher-form symmetries can be used to diagnose intrinsically mixed-state topological order. First, we briefly review the requisite concepts from the study of mixed-states that we will require for our discussion. Following Ref.~\cite{hastings2011finiteT}, throughout we will define a short-range entangled (SRE) mixed state $\rho$ as one which can be expressed as a stochastic mixture of SRE pure-states i.e.,
\begin{align}
    \rho = \sum_k p_k \ket{\psi_k}\bra{\psi_k} \, , \quad \ket{\psi_k} = U_k \ket{0 \dots 0} \, ,
\end{align}
where each state $\ket{\psi_k}$ is SRE i.e., it can be prepared from a product state via a finite-depth unitary circuit $U_k$.

Next, for a mixed-state $\rho$, there are two distinct notions of symmetry~\cite{buca2012,albert2014sym,degroot2022og}: $\rho$ can either have a weak (also called average) or a strong (also called exact) symmetry. $\rho$ is weakly symmetric under a symmetry operator $U$ if $U \rho U^\dagger = \rho$, while $\rho$ obeys a strong symmetry if $U \rho = e^{i \gamma} \rho$ (where $\gamma$ is an overall phase). For a statistical ensemble $\rho = \sum_k p_k \ket{\psi_k} \bra{\psi_k}$ (with $\sum_k p_k = 1$) of pure states $\ket{\psi_k}$, the weak symmetry condition implies that each pure state in the ensemble is symmetric but can carry different symmetry charges ($U \ket{\psi_k} = e^{i \gamma_k} \ket{\psi_k}$), while the strong symmetry condition enforces that each pure state in the ensemble carries the same charge ($U \ket{\psi_k} = e^{i \gamma} \ket{\psi_k} \, \forall k$). Since pure states can only host strong symmetries, intrinsically mixed-states are expected to arise from the non-trivial interplay between strong and weak symmetries. Here, by intrinsically mixed-states we will mean states with some $d$-dimensional quantum order that cannot occur in ground states of gapped local Hamiltonians in $d$ spatial dimensions.

Previous work on the (2+1)D $\Z_2$ Toric Code model has shown how intrinsically mixed-state topological order (imTO) can be characterized via anomalous 1-form symmetries~\cite{Sohal:2024qvq,Ellison:2024svg,Wang2025intrinsic}. Specifically, subjecting a state $\rho$ in the ground-state subspace of this model to a local noise channel $\mathcal{N}$, at maximal decoherence strength, results in the incoherent proliferation of the fermionic particles in $\Z_2$ topological order. The resulting mixed-state $\mathcal{N}\rho$ remains strongly symmetric under the anomalous 1-form symmetry that encodes the fermionic self-statistics of the topological line operator whose end points host the fermionic particles, while the bosonic strong 1-form symmetries of the original pure state are broken to weak 1-form symmetries. The non-modular strong 1-form symmetry of the decohered state underlies its intrinsically mixed nature and, as recently shown in Ref.~\cite{lessa2025higher}, enforces long-range entanglement that places this state in a distinct mixed-state phase of matter than both the original Toric Code phase and the fully trivial phase. This also follows from our discussion in Sec.~\ref{subsec:intrinsic}: as shown in Refs.~\cite{Sohal:2024qvq,Ellison:2024svg,Wang2025intrinsic}, the decohered state $\mathcal{N}\rho$ can be written as a probabilistic mixture of pure-states, each of which is invariant under a strong anomalous 1-form symmetry. Our result thus enforces intrinsic long-range entanglement on the decohered state in the sense that $\mathcal{N}\rho$ has an exponentially vanishing fidelity with any SRE mixed state $\sigma_{\text{SRE}}$. 

Concretely, for generic strong anomalous $p$-form symmetry in generic spatial dimensions, we obtain
\begin{align}
    \mathcal{F}(\mathcal{N}\rho, \sigma_{\text{SRE}}) = O(L^{-\infty})~,
    \label{eq:vanishing fidelity}
\end{align}
if $\mathcal{N}\rho$ stays strongly symmetric.
This is explicitly derived as follows. Since any state $\ket{\Psi}$ in the ensemble of $\mathcal{N}\rho$ is strong symmetric under anomalous $p$-form symmetry, it has a vanishing overlap with any SRE states: $\bra{\Psi}\ket{\text{SRE}}=O(L^{-\infty})$, due to \eqref{eq:small_overlap}. This implies that the trace distance $T(\mathcal{N}\rho,\sigma_{\text{SRE}}):= \frac{1}{2}\text{Tr}\left[\sqrt{(\mathcal{N}\rho-\sigma_{\text{SRE}})((\mathcal{N}\rho)^\dagger-\sigma_{\text{SRE}}^\dagger)}\right]$ becomes
\begin{align}
    T(\mathcal{N}\rho,\sigma_{\text{SRE}}) = 1 - O(L^{-\infty})~.
\end{align}
Since $T(\rho,\sigma) \le \sqrt{1-\mathcal{F}(\rho,\sigma)}$, the fidelity is vanishing: $\mathcal{F}(\mathcal{N}\rho,\sigma_{\text{SRE}})=O(L^{-\infty})$.

Thus, the decohered state $\mathcal{N}\rho$ cannot be obtained from any SRE state via quasi-local finite-depth quantum channels (since these cannot generate long-range entanglement in finite-depth~\cite{zhou2025finiteT,piroli2020}).

We now discuss how our result in Sec.~\ref{subsec:intrinsic} allows us to generalize beyond 1-form symmetries in (2+1)D to obtain a partial characterization of mixed-state topological order via finite $p$-form ($p\geq 1$) symmetries in generic spacetime dimensions. We first present an explicit example in (3+1)D lattice model and then discuss our general conjecture.

\subsection{Noisy (3+1)D Toric Code}
\label{subsec:3dtc}

Since higher-form anomalies enforce intrinsic long-range entanglement in the sense of Eq.~\eqref{eq:small_overlap}, one can straightforwardly see that any strong anomalous $p$-form symmetry leads to a mixed-state where each state in the ensemble has intrinsic long-range entanglement. We now discuss an explicit application of this result to identify a new imTO phase in (3+1)D by subjecting the (3+1)D $\Z_2$ Toric Code to a local noise channel.

Let us consider the (3+1)D $\Z_2$ Toric Code which, for technical reasons, we define here with $\Z_4$ qudits living on each edge $e$ of the 3d cubic lattice:
\begin{align}
    H_{\text{TC}} = -\sum_e X_e^2 -\sum_v (A_v+A^\dagger_v) - \sum_p B^2_p ~,
\end{align}
which can be regarded as starting with the $\Z_4$ Toric Code and then condensing $m^2$ magnetic fluxes to obtain the $\Z_2$ Toric Code. Here, the $A_v$ and $B_p$ terms in the Hamiltonian are the standard vertex and plaquette terms of the (3+1)D Toric Code, defined in terms of the operators $X_e, Z_e$ which satisfy the $\Z_4$ Pauli algebra $Z_e X_e = e^{2\pi i/4} X_e Z_e$.

\begin{figure}[htb]
\centering
\includegraphics[width=0.55\textwidth]{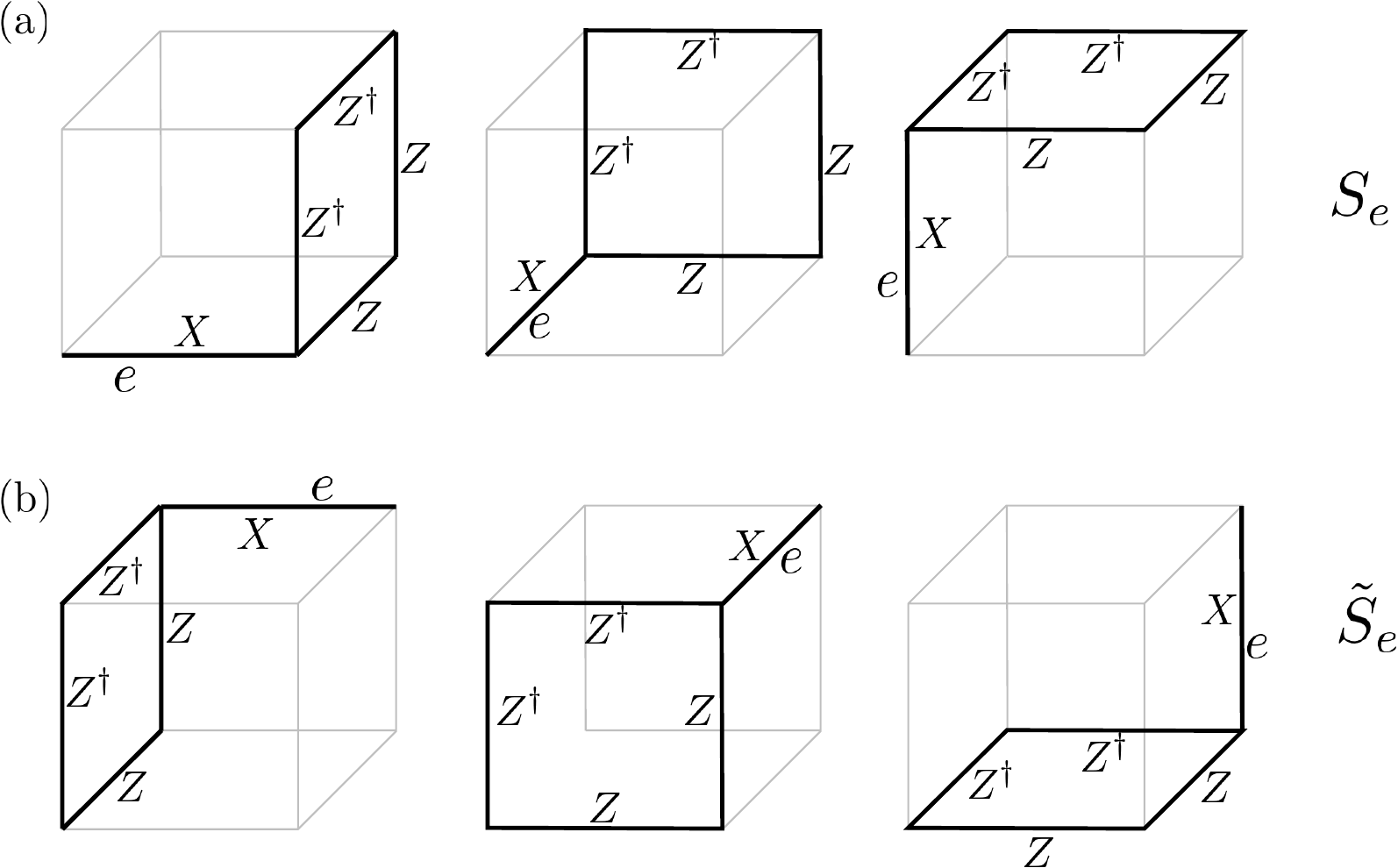}
\caption{Definitions of the operators $S_e, \tilde S_e$.}
\label{fig:Se}
\end{figure}

This $\Z_2$ Toric Code hosts an anomalous $\Z_2$ symmetry, generated by the closed membrane operator
\begin{align}
    S_{\mathrm{f}}(\Sigma) = \prod_{e\subset \Sigma} S_e~,
    \label{eq:floop}
\end{align}
as well as a non-anomalous $\Z_2$ 1-form symmetry generated by
\begin{align}
    S_{\mathrm{b}}(\Sigma) = \prod_{e\subset \Sigma} X_e~,
\end{align}
where $\Sigma$ is a surface on the dual lattice and the product is taken over edges $e$ cut by the surface $\Sigma$. The operator $S_e$ is defined in Figure~\ref{fig:Se} (a).\footnote{
This presentation of fermionic loop creation operators is first discussed in \cite{chen2025toappear}. We thank Yu-An Chen for letting us include it here for a self-contained discussion.} We note that $S_{\mathrm{f}}(\Sigma),S_{\mathrm{b}}(\Sigma)$ define the same operator when $\Sigma$ is closed and oriented since the terms $Z,Z^\dagger$ cancel out. However, these two operators are distinct if $\Sigma$ is either not closed or is closed but non-orientable. The loop excitation corresponding to the anomalous $\Z_2$ 1-form symmetry $S_{\mathrm{f}}$ has the fermionic loop statistics, carrying the nontrivial $\Z_2$ invariant~\eqref{eq: 24-step process} (see Appendix \ref{app:statistics} for the explicit computation of the generalized statistics). We note that defining the (3+1)D $\Z_2$ Toric Code with $\Z_4$ qudits enables us to express the fermionic loops in terms of onsite Pauli operators.

Let us now consider an error channel which leads to the incoherent proliferation of such fermionic loop excitations (in the language of Ref.~\cite{Sohal:2024qvq}, this channel enforces the ``gauging out" of such loop-like excitations):
\begin{align}
    \mathcal{N} = \prod_e\mathcal{N}_e, \quad \mathcal{N}_e(\rho) = p\rho + (1-p) \tilde S_e \rho \tilde S_e^\dagger~,
\end{align}
where $\tilde S_e$ is defined in Figure~\ref{fig:Se} (b). We can faithfully represent the initial density matrix $\rho =\sum_{j,k}\rho_{j,k}\ket{j}\bra{k}$ as a pure-state in a doubled Hilbert space~\cite{choi1975,jamio1972} in terms of the Choi state $\ket{\rho}\rangle = \sum_{j,k}\ket{j}\ket{k}^* \in \mathcal{H}_+\otimes \mathcal{H}_-$, where $\ket{k}^*:= K\ket{k}$ ($K$ denotes complex conjugation) and $\mathcal{H}_+ (\mathcal{H}_-)$ refer to the bra (ket) Hilbert spaces respectively. In the case of maximal decoherence $p=1/2$, the Choi state for the decohered state is expressed as
\begin{align}
    \ket{\mathcal{N}\rho_0}\rangle = \prod_e\left( \frac{1 + \tilde S^+_e (\tilde S^{-}_{e})^*}{2} \right) \ket{\text{TC}}_+\ket{\text{TC}}_-~,
\end{align}
where we took the initial state $\rho_0$ to be a pure ground-state of the (3+1)D Toric Code, $\rho_0 = \ket{\text{TC}}\bra{\text{TC}}$.

The maximally decohered mixed-state is invariant under the strong anomalous $\Z_2$ 1-form symmetry $S_{\text{f}}(\Sigma)$, in the sense that it satisfies the following two properties:
\begin{enumerate}
    \item The closed (oriented) surface operators $S_\text{f}(\Sigma)$ commute with both $\tilde S_{e}^+ (\tilde S^{-}_{e})^*$ and the Toric Code Hamiltonian. Therefore, it remains a strong symmetry of the decohered state:  $S_{\text{f}}(\Sigma) \mathcal{N}\rho_0 =\mathcal{N}\rho_0$.
    \item When the surface $\Sigma$ is not closed, $S_{\text{f}}(\Sigma)$ does not preserve $\mathcal{N}\rho_0$ at the boundary. However, its square commutes with both $\tilde S_{e}^+ (\tilde S^{-}_{e})^*$ and the Toric Code Hamiltonian even when $\Sigma$ has a boundary. Hence, $(S^{+}_{\text{f}}(\Sigma))^2 \mathcal{N}\rho_0 =\mathcal{N}\rho_0$ i.e., this remains a strong symmetry even when $\Sigma$ is not closed\footnote{
    Note that while $S_e^2$ is a product of stabilizers, 
    we also need to ensure that $S_e^2$ commutes with the error channel $\tilde S_{e}^+ (\tilde S^{-}_{e})^*$ to show the invariance of the decohered state $\mathcal{N}\rho$. For instance, $X_e^2$ is also a stabilizer, but it doesn't commute with $\tilde S_{e}^+ (\tilde S^{-}_{e})^*$, so this does not represent a strong symmetry of the decohered mixed-state.}. In this case, $\partial \Sigma$ defines the ``junction'' of $\Z_2$ symmetries where the action of two $\Z_2$ symmetry generators fuses into the vacuum. As explained in Sec.~\ref{subsec:review}, this property of $S_\text{f}(\Sigma)$ is necessary for defining the invariant $U_\Theta$ by the product of $S_{\text{f}}(\Sigma)$. 
\end{enumerate}
Note that while $S_{\text{b}}(\Sigma)$ also satisfies property 1, it violates property 2. The operator $(S_\text{b}(\Sigma))^2$ with open surface $\Sigma$ does not leave $\mathcal{N}\rho_0$ invariant. In other words, requiring that property 2 holds for the $\Z_2$ 1-form symmetry of the decohered density matrix $\mathcal{N}\rho_0$ fixes the boundary of the symmetry operator as $S_{\text{f}}(\Sigma)$ instead of $S_{\text{b}}(\Sigma)$. 

One can then define the invariant $U_\Theta$ using the symmetry operator $S_{\text{f}}(\Sigma)$. The generalized statistics of $S_{\text{f}}(\Sigma)$ for the fermionic loop is shown to be
\begin{align}
    U_\Theta = -1~,
\end{align}
This implies that $\mathcal{N}\rho_0$ is intrinsically long-range entangled, in that each pure-state in the mixed-state ensemble must be an intrinsically LRE state in the sense of Eq.~\eqref{eq:small_overlap}. In particular, the mixed-state $\rho=\mathcal{N}\rho_0$ has exponentially decaying fidelity with any SRE mixed-state in the thermodynamic limit i.e., it satisfies Eq.~\eqref{eq:vanishing fidelity}.

Given this result, it is straightforward to show that the decohered state $\mathcal{N} \rho_0$ belongs to a distinct mixed-state phase of matter than both the initial (3+1)D Toric Code ground state and the fully trivial phase. Moreover, it constitutes a new intrinsically mixed-state topological order in (3+1)D, akin to how the fermionic decohered Toric Code in (2+1)D represents an imTO phase due to its strong anomalous 1-form symmetry~\cite{Sohal:2024qvq,Ellison:2024svg,zhang2024,lessa2025higher}. First, note that the decohered state $\mathcal{N} \rho_0$ carries no logical information since the state only weakly preserves the $\Z_2$ 1-form symmetry generated by the magnetic flux, meaning that the logical Pauli $X$ gate becomes a weak symmetry. Thus, we can no longer use this operator to distinguish between different logical states (equivalently, we no longer have a representation of the Pauli algebra on the code-space)\footnote{This is similar to the case of the (3+1)D Toric Code at small but finite temperature, where the particle excitations proliferate while the loop excitations do not---it is well-known that this intermediate regime constitutes a classical topological memory~\cite{chamon2007}.} Hence, there can not exist any recovery channel $\mathcal{R}$ that restores the decohered state back to the initial pure-state, which encodes nontrivial logical information i.e., $\nexists \mathcal{R}: \mathcal{R}\mathcal{N}\rho_0=\rho_0$. Note that while the decohered state does not support a quantum memory, it still hosts a classical memory due to the remaining strong anomalous fermionic symmetry (similarly to the so-called $f$-decohered Toric Code in (2+1)D).

Thus, according to the two-way channel equivalence relation on mixed-state phases defined in Ref.~\cite{sang2023mixed}\footnote{According to this definition, two states $\rho_1$ and $\rho_2$ belong to the same mixed-state phase of matter iff $\exists \, \Sigma_{12}, \Sigma_{21}: \Sigma_{12} \rho_2 = \rho_1 \& \Sigma_{21} \rho_1 = \rho_2$, where $\Sigma_{12}, \Sigma_{21}$ are quasi-local quantum channels whose depth scales at most polylogarithmically with system size. Note that this equivalence relation is coarser than one defined in terms of the Markov length of the conditional mutual information~\cite{sang2024}; showing phase inequivalence of two states under the former thus automatically implies phase inequivalence under the latter.}, $\rho_0$ and $\mathcal{N} \rho_0$ must belong to distinct mixed-state phases of matter. Next, $\mathcal{N} \rho_0$ cannot be prepared from any SRE state via a finite-depth quasi-local channel due to its intrinsic long-range entanglement (neither such channels nor stochastic local channels can produce long-range entanglement~\cite{zhou2025finiteT,piroli2020,lessa2025higher} in finite-depth since they respect Lieb-Robinson bounds): thus, the decohered state is also in a distinct mixed-state phase of matter than the completely trivial phase. This shows that this state hosts a distinct mixed-state topological order in (3+1)D. We now argue that this form of topological order is in fact intrinsic to mixed-states in that it is not expected to arise in the ground states of gapped local Hamiltonians.

\subsubsection{Intrinsically mixed-state topological order}
The surface operators for the fermionic loop generate an anomalous one-form symmetry of (3+1)D theory. The anomaly can be captured on the lattice as a topological index given by the Berry phase for a 36-step process for the membrane operators that generate the one-form symmetry \cite{Fidkowski:2021unr,Kobayashi:2024dqj}.
Another to describe the anomaly is using the bulk-boundary correspondence, where the bulk has an SPT with the response action~\cite{Kapustin:2017jrc,Fidkowski:2021unr,Chen:2021xks,kobayashi2024universal}:
\begin{equation}
\pi\int B\cup \frac{d\hat{B}}{2}=    \pi\int w_3\cup B~,
\end{equation}
where $B\in Z^2(M^5,\Z_2)$ is the background $\mathbb{Z}_2$ two-form gauge field for the one-form symmetry, and $w_3$ is the third Stiefel-Whitney class. $\hat{B}$ is the $\Z_4$ lift of the $\Z_2$ gauge field, so that $d\hat{B}$ takes value in 0,2 mod 4. The above topological response describes a nontrivial SPT phase with $\Z_2$ 1-form symmetry in (4+1)D. 

In a pure state defined on a tensor product Hilbert space, realizing an anomalous one-form symmetry within a gapped phase necessarily requires nontrivial topological order. In particular, in a pure-state topological order in (3+1)D, every loop excitation is accompanied by a particle excitation that exhibits nontrivial mutual braiding with it.

Meanwhile, we have explicitly shown above that the mixed-state obtained by maximal decoherence of the fermionic loop excitations in the (3+1)D Toric Code results in a topological order where the only remaining strong symmetry is the anomalous surface operators, without any line operators and corresponding particle excitations. 
Unlike the original pure state, where this 1-form symmetry had mixed anomalies with other 2-form symmetries (electric particles) as captured by the braiding process in the remote detection,
the decohered mixed-state does not host any other strong 1-form symmetries and is not detectable by braiding with other topological operators that leave the state invariant. Such forms of topological order, where the notion of remote detectability breaks down, are widely believed to not occur in the ground states of gapped local Hamiltonians with tensor product Hilbert space. Thus, this mixed-state represents an imTO phase in (3+1)D.

Similar to  the (2+1)D case discussed in Refs.~\cite{Sohal:2024qvq,Ellison:2024svg}, we can view our imTO as obtained from a (4+1)D bulk $\mathbb{Z}_2$ two-form gauge theory, specified via the topological action
\begin{equation}
    \pi\int B\cup \frac{d\hat{B}}{2}~,
\end{equation}
where the fermionic loop excitation on the boundary is the magnetic loop excitation of the two-form gauge field $B$. The bulk theory has nontrivial topological order and is similar to the (3+1)D fermionic $\mathbb{Z}_2$ gauge theory that describes the boundary transparent fermion. Heuristically, our intrinsically mixed-state can be obtained by putting this bulk theory on a slab geometry and tracing out one of the boundaries.

\subsection{Implications for Mixed-State Topological Orders}
\label{subsec:gen}

We close with a brief discussion of the general implications of our result on mixed-state topological orders in arbitrary dimensions. An important aspect of the generalized statistics is that they are well-defined for any symmetric state~\cite{kobayashi2024universal} without any additional details on the state or Hamiltonian. Therefore, our constraint is applied to any mixed-states with strong anomalous higher-form symmetries, and such mixed-states must be intrinsically long-range entangled.

Therefore, there is a clear route to generalizing new example of imTO phase in (3+1)D in arbitrary dimensions via the algebraic structure of their strong higher-form symmetries. First, we can consider other fixed-point pure-states with topological order in $d$-spatial dimensions and subject them to maximal decoherence that induces the incoherent proliferation of extended excitations with nontrivial statistics, such as the ``fermionic membranes'' systemically classified in Ref.~\cite{kobayashi2024universal}. As laid out in Ref.~\cite{Sohal:2024qvq}, excitations with arbitrary self-statistics can be incoherently proliferated (or ``gauged out"): this is striking in contrast with the coherent condensation of excitations in pure state topological orders, where only bosonic excitations are permitted to condense. Physically, this occurs due to a discrete gravitational framing anomaly for the symmetries corresponding to excitations with nontrivial self-statistics, such that their coherent proliferation would give rise to nontrivial ``transparent" excitations that violate remote detectability. Since there is no obstruction to the incoherent proliferation of such excitations via finite-depth quasi-local quantum channels, mixed-states can realize more general topological orders which violate remote detectability. In particular, the algebraic structure of the strong higher-form symmetries characterizing mixed-state topological order can violate remote detectability as we have already shown by explicit example in (3+1)D. Thus, it is reasonable to conjecture that the breakdown of remote detectability as encoded in the algebra of strong anomalous higher-form symmetries is a sufficient criteria for distinguishing intrinsically mixed-state topological order from pure-state topological order in arbitrary spatial dimensions. Indeed, in forthcoming work~\cite{3dimto}, we will provide a classification of imTO in (3+1)D in terms of braided fusion 2-categories.

Coupled with our constraint that strong anomalous higher-form symmetries enforce intrinsic long-range entanglement, this also means that the mixed-states with non-detectable anomalous strong higher-form symmetries obtained via local decoherence of fixed-point pure-states belong to genuine imTO phases of matter, as the intrinsic LRE condition implies that such states cannot be two-way channel connected to either the original pure-state or a trivial SRE mixed-state. We note that we are entirely ignoring the algebraic structure of weak higher-form symmetries and their interplay with the strong higher-form symmetries: as discussed in Ref.~\cite{Sohal:2024qvq}, this structure is crucial for understanding the set of locally detectable quasiparticles (which generalize the notion of excitations above a gapped ground state). We leave a thorough investigation of the weak symmetries of imTO phases to future work. 

%%%%%%%%%%%%%%%%%%%%%%%%%%%%%%%%%%%%%%%%%%%%%%%%%%%%%
%%%%%%%%%%%%%%%%%%%%%%%%%%%%%%%%%%%%%%%%%%%%%%%%%%%%%

\section{Conclusions}
\label{sec:cncls}

In this paper, we have demonstrated that finite anomalous higher-form symmetries enforce intrinsic long-range entanglement on quantum many-body states. As a direct consequence, the strong anomalous higher-form symmetries enforce intrinsic long-range entanglement on mixed-states obtained via locally decohering pure ground states, with the anomaly enforcing an exponentially decaying fidelity between the mixed-state and any short-range entangled state. As an application, we obtained a new instance of an intrinsically mixed-state topological order in (3+1)D by considering the $\Z_2$ Toric Code subject to decoherence, with the resulting decohered state characterized by a strong anomalous $\Z_2$ 1-form symmetry that violates remote detectability. 

Our work opens up a number of intriguing directions. First, we have focused entirely on invertible higher-form symmetries in this paper: what are the constraints on entanglement stemming from non-invertible higher-form symmetries? Second, we have conjectured that topological order that violates remote detectability condition can generally be realized as the algebraic structure of strong symmetries of mixed-states. It would be interesting to further identify other criteria in pure-state topological orders whose violation leads to intrinsically mixed-state topological order and to pursue a full characterization of such mixed-state phases. It also remains an open question to what extent the algebraic framework for understanding gapped boundaries or gapped domain walls in pure-state topological orders generalizes to mixed-state topological orders. In the pure state context, gapped boundaries of (2+1)D topological orders are well understood and are characterized by so-called Lagrangian algebras; in contrast, the algebraic data for identifying analogues of gapped boundaries of even the simplest imTOs in (2+1)D remains far from clear. Analogously, it would be of interest to understand the structure of the strong/weak symmetries of the gapped boundaries, which would correspond to boundary topological operators. Finally, crisp diagnostics of mixed-state topological orders beyond entanglement probes (such as the topological entanglement negativity) should be developed, with a view towards experiments on NISQ platforms, where circuit level noise rapidly decoheres carefully prepared topologically ordered pure-states. For instance, the Fredenhagen-Marcu order parameter~\cite{fm83,fm86,fm88,Gregor:2010ym} for detecting ground state topological order has been experimentally measured~\cite{semeghini2021} and it would be of interest to obtain an analogous order parameter for diagnosing mixed-state topological order.

%%%%%%%%%%%%%%%%%%%%%%%%%%%%%%%%%%%%%%%%%%%%%%%%%%%%%
%%%%%%%%%%%%%%%%%%%%%%%%%%%%%%%%%%%%%%%%%%%%%%%%%%%%%

\section*{Acknowledgments}
We thank Yu-An Chen, Zhu-Xi Luo, Ramanjit Sohal, Beni Yoshida, Matthew Yu, and Carolyn Zhang for stimulating discussions. R.K. and A.P. are supported by the U.S. Department of Energy, Office of Science, Office of High Energy Physics under Award Number DE-SC0009988 and by the Sivian Fund at the Institute for Advanced Study. A.P. acknowledges support from the Paul Dirac Fund at the Institute for Advanced Study. P-S.H. is supported by Department of Mathematics, King's College London. R.K. and P.-S.H. thank the Kavli Institute for Theoretical Physics for hosting the program “Generalized Symmetries in Quantum Field Theory: High Energy Physics, Condensed Matter, and Quantum Gravity” in 2025, during which part of this work was completed. This research was supported in part by grant no. NSF PHY-2309135 to the Kavli Institute for Theoretical Physics (KITP). The authors of this paper were ordered alphabetically.

%%%%%%%%%%%%%%%%%%%%%%%%%%%%%%%%%%%%%%%%%%%%%%%%%%%%%
%%%%%%%%%%%%%%%%%%%%%%%%%%%%%%%%%%%%%%%%%%%%%%%%%%%%%

\appendix

%%%%%%%%%%%%%%%%%%%%%%%%%%%%%%%%%%%%%%%%%%%%%%%%%%%%%
%%%%%%%%%%%%%%%%%%%%%%%%%%%%%%%%%%%%%%%%%%%%%%%%%%%%%

\section{Review: Simplicial Complex}
\label{app:complex}

In this Appendix, we briefly review the notions of chain complexes used in the main text, focusing on the definition of the $(d-p)$-chain $C_{d-p}(X,G)$ and its boundary $B_{d-p-1}(X,G)$ on a simplicial complex $X$ with coefficient group $G$.
Let $X$ be a triangulation (simplicial complex) of a $d$-dimensional spatial manifold. Each oriented $k$-simplex is denoted by an ordered set of vertices $\langle i_0 i_1 \cdots i_k\rangle$, with $i_0 < i_1<\dots < i_k$. The group of $k$-chains with coefficients in an Abelian group $G$ is defined as the direct sum
\begin{align}
C_k(X,G) = \bigoplus_{\sigma_k \in X_k} G[\sigma_k],
\end{align}
where $X_k$ is the set of oriented $k$-simplices of $X$. Elements of $C_k(X,G)$ are formal sums
\begin{align}
c = \sum_{\sigma_k} g_{\sigma_k} [\sigma_k], \qquad g_{\sigma_k} \in G,
\end{align}
so that each simplex $\sigma_k$ is labeled by an element $g_{\sigma_k}$ of $G$. In the present context, the symmetry operator $U_g(\Sigma)$ supported on a $(d-p)$-dimensional manifold $\Sigma$ is represented by a chain
\begin{align}
s = \sum_{\sigma_{d-p}} g_{\sigma_{d-p}} [\sigma_{d-p}] \in C_{d-p}(X,G),
\end{align}
which specifies how the local symmetry actions are distributed over the $(d-p)$-simplices of $X$.
The boundary operator $\partial_k: C_k(X,G) \to C_{k-1}(X,G)$ is defined on an oriented simplex by
\begin{align}
\partial_k \langle i_0 i_1 \cdots i_k\rangle = \sum_{m=0}^k (-1)^m \langle i_0 \cdots \hat{i}_m \cdots i_k\rangle,
\end{align}
where $\hat{i}_m$ means that the vertex $i_m$ is omitted. Extending linearly to all chains, this operator satisfies the fundamental property $\partial_{k-1} \circ \partial_k = 0$. The image of $\partial{k+1}$ defines the group of boundaries, $B_k(X,G) = \mathrm{Im}(\partial_{k+1})$, while its kernel defines the group of cycles, $Z_k(X,G) = \ker(\partial_k)$, with $B_k(X,G) \subseteq Z_k(X,G)$.
In the main text, the open symmetry operator labeled by $s \in C_{d-p}(X,G)$ creates a configuration of symmetry defects on its boundary, described by $a = \partial s \in B_{d-p-1}(X,G)$. Physically, the element $a$ corresponds to the locus where the higher-form symmetry operator terminates, and the resulting defects are located on the $(d-p-1)$-simplices that form the boundary of the support of $s$. Each such configuration $a$ defines a state $|a\rangle$ obtained by acting with a suitable combination of open symmetry operators on the symmetric state $|\Psi\rangle$, as discussed in Eq.~\eqref{eq:def of theta} of the main text.

 The generalized statistics invariant introduced in Eq.~(3) of the main text can then be viewed as the Berry phase obtained from an ordered product of symmetry operators $U(s_j)$ (or its inverse) labeled by chains $s_j \in C_{d-p}(X,G)$, whose boundaries $\partial s_j \in B_{d-p-1}(X,G)$ encode the locations of the created symmetry defects. The resulting phase $e^{i\Theta}$ captures the obstruction to gauging the finite $p$-form symmetry, and hence provides a microscopic definition of the ’t~Hooft anomaly in lattice models.

\section{Computation of fermionic loop statistics}
\label{app:statistics}

\begin{figure}[b]
\centering
\includegraphics[width=0.5\textwidth]{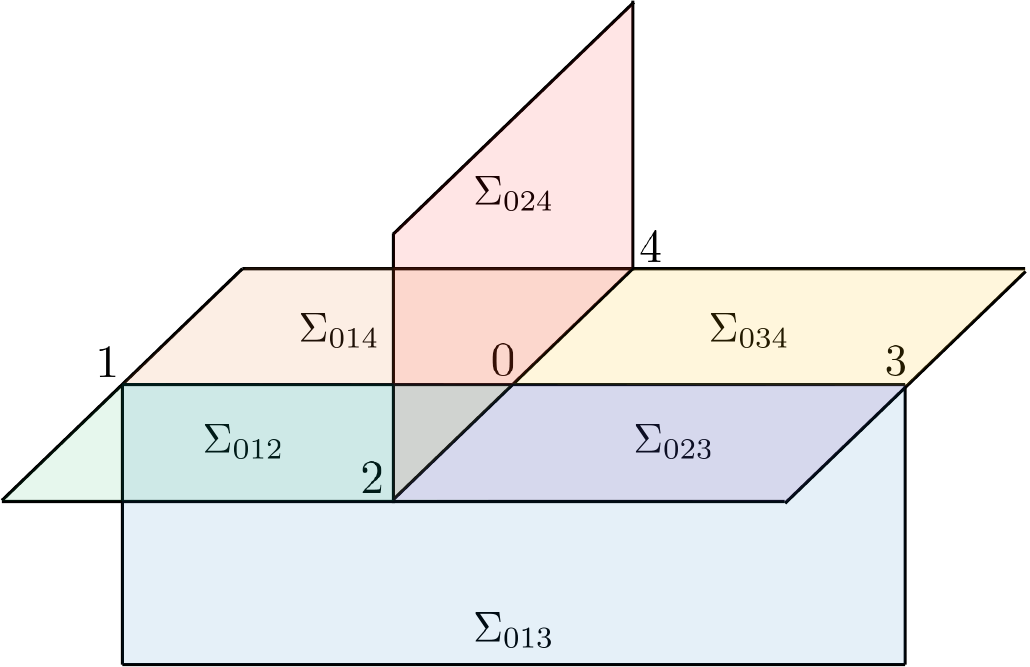}
\caption{The configurations of surfaces $\Sigma_{0jk}$ embedded in the dual cubic lattice.}
\label{fig:Sigma}
\end{figure}

In this Appendix, we compute the generalized statistics of the $\Z_2$ 1-form symmetry operator $S_{\text{f}}$ introduced in Eq.~\eqref{eq:floop}. We consider the symmetry operators $U_{0jk}:= S_{\text{f}}(\Sigma_{0jk})$ with open surfaces $\Sigma_{0jk}$, as shown in Figure \ref{fig:Sigma}. We compute the generalized statistics 
    \begin{eqs}
        U_\Theta := ~& U_{014} U_{034} U_{023} U_{014}^{-1} U_{024}^{-1} U_{012} U_{023}^{-1} U_{013}^{-1} \\
        \times & U_{024} U_{014} U_{013} U_{024}^{-1} U_{034}^{-1} U_{023} U_{013}^{-1} U_{012}^{-1} \\
        \times & U_{034} U_{024} U_{012} U_{034}^{-1} U_{014}^{-1} U_{013} U_{012}^{-1} U_{023}^{-1}~.
    \end{eqs}
One can simplify the above sequence of operators using the following two properties:
\begin{itemize}
\item Due to our choice of operators, $U_{012}, U_{023}, U_{034}, U_{014}$ commute with each other.
\item Since $\{U_{0jk}\}$ are Pauli operators, the commutator in the form of $[U,U']:=UU'U^{-1} U'^{-1}$ with any subscripts becomes a phase. In particular, this allows us to rewrite the sequence as
\begin{align}
    ... UU'U^{-1} ... = ... U' ... \times [U,U']~.
\end{align}
\end{itemize}
By utilizing the above properties iteratively, we arrive at the expression
\begin{align}
\begin{split}
    U_\Theta =[U_{024},U_{013}]^2 [U_{023}, U_{024}][U^{-1}_{023}, U_{024}][U_{012}, U_{024}][U_{012}^{-1}, U_{024}][U_{034},U_{013}][U^{-1}_{034},U_{013}][U_{013},U_{023}][U^{-1}_{013},U_{023}]
    \end{split}~.
\end{align}
The above commutators, except for the first one, cancel out by $[U,U'][U^{-1},U']=1$. We then get 
\begin{align}
\begin{split}
    U_\Theta =[U_{024},U_{013}]^2~.
    \end{split}
\end{align}
In the above expression, $U_{024},U_{013}$ shares a single $\Z_4$ qudit where $U_{024}$ has $X$ and $U_{013}$ has $Z^\dagger$. Therefore, we obtain $U_\Theta = -1$, implying that $S_{\text{f}}$ has fermionic loop statistics.

%%%%%%%%%%%%%%%%%%%%%%%%%%%%%%%%%%%%%%%%%%%%%%%%%%%%%
%%%%%%%%%%%%%%%%%%%%%%%%%%%%%%%%%%%%%%%%%%%%%%%%%%%%%

\section{Weak fermionic loop symmetry and entanglement entropy}
\label{app:average}

\begin{figure}[t]
    \centering
    \includegraphics[width=0.5\linewidth]{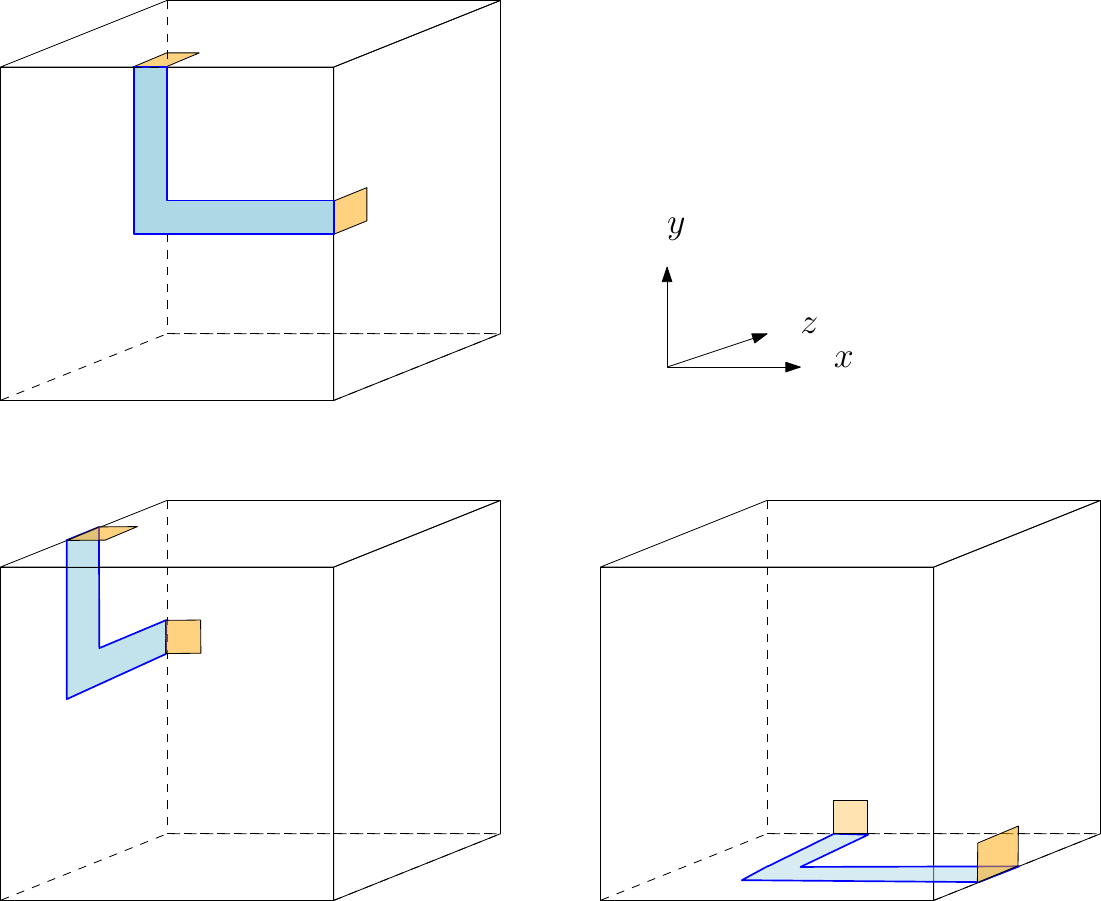}
    \caption{Fermionic loop symmetry and entangling surface. Each loop denotes the boundary of an open membrane operator inside the box for the fermionic loop. Since everything outside the box is traced out, near the blue membrane the loops are created by $\prod Z$.
    We can label the blue loops using a pair of orange loops, such that the membrane operator for the blue loop does not commute with the membrane operator for each of the two orange loops due the commutator between $X,Z$. The pairs of orange loops correspond to the set of lattice coordinate points on the entire plane where the blue loop is located.
}
    \label{fig:entanglement}
\end{figure}

Let us consider the implication of fermionic loop symmetry on the entanglement entropy using the method developed in Ref.~\cite{Hsin:2023jqa}. Consider the entanglement entropy for the ball-shape region $R$ enclosed by the entangling surface, as shown in Fig.~\ref{fig:entanglement}. For any state that has the strong fermionic loop symmetry, it also has the weak (or average) fermionic loop symmetry. Since the fermionic loop symmetry is a depth-1 circuit, the reduced density matrix $\rho_R$ also has the weak fermionic loop symmetry~\cite{Hsin:2023jqa}. The commutation relation for the symmetry then gives a bound on the entanglement entropy for the region, given by the minimal dimension $d_M$ of the project representation that realizes the commutation relation
\begin{equation}
    S(\rho_R)\geq \log d_M~.
\end{equation}

To find the dimension $d_M$, we can construct the representation using an auxiliary Fock vacuum $|0\rangle$ and applying the symmetry operators.
The reduced fermionic loop symmetry is generated by both closed membranes within the region and open membranes terminating on the entangling surface, where the other part outside the region is traced out. Let us apply the open membrane operator whose boundary is the blue loop shown in Fig.~\ref{fig:entanglement}, which can be labeled by a lattice coordinate on each face. The states obtained from $|0\rangle$ by applying different blue fermionic loop operators can be distinguished via the commutator with the pair of membrane operators whose boundaries are the pair of orange fermionic loops near the coordinate axes (see Fig.~\ref{fig:entanglement}). This gives at least $d_M\geq 4^{L_xL_y+L_yL_z+L_xL_z}=4^{A/2}$ where $L_x,L_y,L_z$ are the lengths of the box in the $x,y,z$ directions, and $A$ is the total area of the entangling surface (i.e. the surface of the box). Thus, the entanglement entropy is bounded by
\begin{equation}
    S(\rho_A)\geq \log d_M\geq A\log 2~.
\end{equation}
Thus, any mixed-state that is invariant under the weak fermionic loop symmetry has a non-vanishing entanglement entropy. In the future, it would also be interesting to compute the entanglement negativity for states that are either strongly or weakly symmetric under the fermionic loop symmetry.

%%%%%%%%%%%%%%%%%%%%%%%%%%%%%%%%%%%%%%%%%%%%%%%%%%%%%
%%%%%%%%%%%%%%%%%%%%%%%%%%%%%%%%%%%%%%%%%%%%%%%%%%%%%

\bibliographystyle{utphys}
\bibliography{bibliography}

%%%%%%%%%%%%%%%%%%%%%%%%%%%%%%%%%%%%%%%%%%%%%%%%%%%%%
%%%%%%%%%%%%%%%%%%%%%%%%%%%%%%%%%%%%%%%%%%%%%%%%%%%%%

\end{document}